\documentclass[a4paper]{jpconf}
\usepackage{graphicx}
\usepackage{physics}

\def\d{{\rm d}}

\def\7{$\;$}
\def\l{\left}
\def\r{\right}
\def\be{\begin{equation}}
\def\ee{\end{equation}}
\def\bea{\begin{eqnarray}}
\def\eea{\end{eqnarray}}
\def\f{\frac}

\begin{document}
\title{Nature of thermodynamics equation of state towards economics equation of state}

\author{Burin Gumjudpai}
\address{The Institute for Fundamental Study ``The Tah Poe Academia Institute", \\ Naresuan University, Phitsanulok 65000, Thailand}
\address{Thailand Center of Excellence in Physics, Ministry of Education, Bangkok 10400, Thailand}
\address{Graduate School of Development Economics, \\
National Institute of Development Administration (NIDA),
Bangkok 10240, Thailand}
\ead{buring@nu.ac.th}
\author{Yuthana Sethapramote}
\address{Graduate School of Development Economics, \\
National Institute of Development Administration (NIDA),
Bangkok 10240, Thailand}


\begin{abstract}
This work critics on nature of thermodynamics coordinates and on roles of the variables in the equation of state (EoS). Coordinate variables in the EoS are analyzed so that central concepts are noticed and are used to lay a foundation in building of a new EoS or in testing EoS status of a newly constructed empirical equation.  With these concepts, we classify EoS into two classes. We find that the EoS of market with unitary price demand and linear price-dependent supply function proposed by \cite{GumjMarket}, is not an EoS because it has only one degree of freedom.
\end{abstract}

\section{Introduction}
\label{Section:Introduction}
There are two classes of physics, the first are those derived from least action principle and the second is the equilibrium thermodynamics (or thermostatics). Both emerge from either postulated setting or empirical results. Empirical laws in thermostatics are usually the equation of state (EoS) - a mathematical function relating temperature ($T$), intensive coordinate ($Y$) and extensive coordinate ($X$) together in from of $g(X, Y, T) = 0$. In a coordinate space of $(X, Y, T)$, the EoS is a constraint surface reducing one degree of freedom out of the three, leaving with only two degrees of freedom on the surface. There are some significant aspects of these variables needed to be considered. These aspects are the roles of thermodynamics coordinate variables and how the variables affect each other. We analyze how well-known EoS are expressed so that we can comment and justify the EoS status of an empirical relation. Particularly in this work, we try justifying the economics EoS proposed previously in Gumjudpai (2018) \cite{GumjMarket}. Consideration of existence of the economics EoS could lay out some novel approach to economics employing thermodynamics. A few proposal on thermodynamical economics previously are, for examples, the works of
E. Smith and D. K. Foley (2008)  \cite{SF} and Saslow (1999) \cite{saslow} with different ways of connecting economics to thermodynamics. In E. Smith and D. K. Foley (2008)
\cite{SF},  utility is analogous to entropy and mechanical pair are commodity
(extensive coordinate) and marginal rate of substitution (intensive coordinate). The economics version of the first law
of thermodynamics is the conservation of commodities (instead of the internal energy). Utility is maximized as entropy.
In Saslow (1999) \cite{saslow}, utility function is analogous to internal
energy, wealth to the Helmholtz free energy, price to chemical potential, quantities of
commodity to particle number, economic surplus quantity to $TS$ (multiplication of temperature and entropy). In Gumjudpai (2018) \cite{GumjMarket}, motivated by Carath\'{e}odory's axiom (see, e.g. Buchdahl (1966) \cite{buchdahl} and M\"{u}nster (1970) \cite{munster}), construction of EoS empirically is preferred as initial step for deduction rather than beginning with finding the maximized variable. Potentials (e.g. internal energy) are defined as consequence of existence of the EoS. We follow the Carath\'{e}odory's axiom, therefore we look at building blocks of the EoS in detail and hope to lay a step towards the complete theory of thermodynamics formulation of economics. Here we study nature of thermodynamics variables and see how they relate. We propose a diagram describing patterns of effect among the EoS variables and we try to find some criteria for naturalness of an EoS. At last, we check the naturalness of economics EoS proposed previously in Gumjudpai (2018) \cite{GumjMarket}.

\section{Roles of Thermodynamics Coordinate Variables in an EoS}
\subsection{Endogenous and Exogenous Effects}
Endogenous and exogenous variables are common economics terms. Consider set of real-value variables, $\{x_1, \ldots , x_n \}$ and let there be a function,
\be
x_1  =  f(x_2, \ldots, x_i, \ldots x_n) \,.
\ee
This function $f$ is made or assumed when one tries to deal with data obtained from observation or surveying. 
In many cases, doing surveying or observation is performed without concrete model of the system. 
Data are collected in scattered way and variables can not be fully controlled. This is unlike experimental data of which we typically have model in hand to be tested. Controlled group and controlled variables can be arranged or set.  Endogenous variables are the variables thought by researchers to be included in the model whereas exogenous variables are variables considered as external factors of the model.
However, we do not know really whether the considered model is correct if without prediction power together with good theoretical background.   In experimental physics, we usually have concrete model given by postulate laws thus endogenous and exogenous variables are more or less known prior to performing experiment or conducting observation. In addition, some variables can be controlled.  This is where deduction meets induction reasoning.  In social sciences, the model and postulate are not solid and data collection is usually performed with surveying. Therefore it is not clearly known which variables are endogenous or exogenous and one has to deal with scattered data obtained from surveying. Therefore in building a model, one has to set some variables to be endogenous according to his or her assumption and concepts.
If $(x_2, \dots, x_i)$ are taken to be endogenous variables of $f$ for constant values of all other variables, i.e. $(x_{i+1}=a_{i+1}),\ldots, (x_n=a_n)$, that is
 \be
x_1  = f(x_2, \ldots, x_i, a_{i+1}, \ldots, a_n) = f_1(x_2, \ldots, x_i), 
\ee
hence we say that $(x_2, \dots, x_i)$ have endogenous effect on $x_1$. If $(a_{i+1},\ldots, a_n)$ change their values, these variables $(x_{i+1},\ldots, x_n)$ are said to be exogenous variables and to have exogenous effects on $x_1$.  This is with a basis of thought that the factors of change that affect $x_1$ in the model are only
$x_2, \ldots, x_i$ which are considered as the only set of  independent variables in framework of the model.  When doing regression analysis,  all other $(a_{i+1}, \ldots, a_n)$ are regarded as constant of the regression. If we change conceptual framework of the model to include some exogenous variables to the model, for example, including of $x_{i+1}$ to the model, hence  $x_{i+1}$ becomes endogenous and 
 \be
x_1  = f(x_2, \ldots, x_i, x_{i+1}, a_{i+2}, \ldots, a_n) = f_2(x_2, \ldots, x_{i+1}). \ee
In economics, exogenous change is sometimes known as shock of demand or supply functions in markets of different types of product, e.g. labours, loanable funds, etc.

\subsection{Role of Variables}
Thermodynamics coordinates are not equivalent although each of them is axis of 3-dimensional thermodynamics space. The three variables are $T$, $Y$ (intensive coordinate) and $X$ (extensive coordinate). Temperature is a fundamental thermodynamics coordinate since it is laid by the 0th law of thermodynamics. Pair variables $(X,Y)$ is dubbed the mechanical pair because they are from the type of physics derived from Hamilton's principle. Mechanical pair must form a work term, $\delta W = Y \d X$. 

\subsubsection{Examples of Mechanical Pair:}
\begin{itemize}
  \item In hydrostatics system, $X$ is the volume $V$, $Y$ is $-P$ (minus sign of pressure).
  \item In paramagnetics substance, $X$ is the magnetization $\mathcal{M}$, $Y$ is an external magnetic field intensity $B_0$.
  \item In one-dimension elastic string,   $X$ is $\ell \equiv x^2/2$ and $Y$ is $ f \equiv F/x $ (force per length).
  \item In a two-surface rectangular thin film with length $x$ and with constant width $l$, $X$ is area $A = lx$ and $Y$ is $2 \Gamma $ where $\Gamma = F/l$ is the surface tension.
  \item In a reversible electrolytic cell, $X$ is charge $q$ and $Y$ is electromotive force $\mathcal{E}$ which is an electrical energy per unit of charge.
  \item    In dielectric substance, $X$ is total electric dipole moment $\mathcal{P}$ and $Y$ is an external electric field intensity $E$.
\end{itemize}
\subsubsection{Active and Passive Roles of Extensive Coordinate, $X$:}
Considering only endogenous effect, in some cases, $X$ has a passive role, i.e. it takes an effect of other influence that causes changing in $X$. For instance, the volume $V$ can not change without the effect of pressure $P$. In paramagnetics substance, magnetization $\mathcal{M}$ can not change unless being affected by magnetic field intensity, $B_0$ or $T$. In this case $X$ (that is $\mathcal{M}$) can take active role, i.e. changing in $\mathcal{M}$ (as an effect of $B_0$) can cause changing in $T$.  The cause of change in $T$ this way come originally from $B_0$ but via $\mathcal{M}$. 

\subsubsection{Active and Passive Roles of Intensive Coordinate, $Y$:}
Intensive variables $Y$ are physical quantities with nature of {\it influence}, e.g. 
\begin{itemize}
  \item (1) force, power or energy {\it per unit of the quantity}  which the influence has an endogenous effect on.
  Example of $Y$, that is defined with influence per unit of quantity $X$ it has effect on, is the electromotive force $\mathcal{E}$ in reversible (ideal) electrolytic cell. $\mathcal{E}$ is energy per unit charge ($X$ is charge $q$) and $\mathcal{E}$ causes motion of charge quantities $q$.
  \item (2) force, power or energy {\it per unit of related quantity} which the influence has an endogenous effect on. $Y$ is an influence per unit of a quantity related to $X$ (instead of per unit of $X$). For examples, in hydrostatics system, $Y$ is $-P$ which is force per area, $-F/A$. The extensive $X$ is volume $V$. Volume $V$ is related to area $A$. In elastic string, $Y$ is $f = F/x$ but $X$ is $\ell \equiv x^2/2$. In two-surface thin film, $Y \equiv 2\Gamma = 2 F/l$ and $X \equiv A = lx$. In case of paramagnetics, $Y$ is $B_0$ which has a spirit of force per magnetic charge (not exist in reality) which is related to the extensive $\mathcal{M}$. In dielectrics, $Y$ is electric field intensity $E$ which is a force per unit charge. Charge is related to $X$ that the total electric dipole moment $\mathcal{P}$ (which is $X$) is $\vec{\mathcal{P}} = \sum q \vec{d_i} $ where $\vec{d_i}$ points from negative to positive charge.
\end{itemize}
Although $Y$ has active role as influence, it also has passive role, i.e. $T$ can cause endogenous changes in $-P$ in a hydrostatic system.

\subsubsection{Active and Passive Roles of  Temperature, $T$:}
  $T$ identifies thermal state of a system. $T$ must have both active and passive roles. In hydrostatic system, $T$ can endogenously affect $Y$ (that is $-P$) and reversely $P$ can endogenously affect $T$. In paramagnetics system, $T$ can endogenously affect  $\mathcal{M}$ and reversely $\mathcal{M}$ can endogenously affect $T$.


\section{Truly Endogenous Function}
In describing accurately and concisely a function that encodes direct cause and direct effect of independent and dependent variables, we must not include cause and effect via composite function.
We need a concrete expression for a direct relation between variables so that we have tool for deeper intuition in describing a system.

In perfect paramagnetics substance, considering that $\mathcal{M}$ has direct effect on $T$.  $B_0$ affects $T$ indirectly via $\mathcal{M}$. Thus $B_0$ is exogenous variable of $T$. However the EoS (Curie's law) is written as
\be
\mathcal{M} =  C\f{B_0}{T} \;\;\;\;\;\;\;   \text{or} \;\;\;\;\;\;\;  T   =  C\f{B_0}{\mathcal{M}}
\ee
that is 
\be
T =  T(\mathcal{M},B_0).
\ee
As we realize that $B_0$ affects $T$ via $\mathcal{M}$, thus this is in fact,
 \be 
 T = T(\mathcal{M}(B_0)) =  T\circ\mathcal{M}(B_0)\,. 
 \ee
In ideal gas system, we know that $V$ depends directly only on $P$. The reverse is not true in reality. $P$ is the cause in changing of $V$. Volume can not change its value if the pressure does not push it.  
Temperature $T$ can not affect $V$ directly but via the pressure $P$, i.e. temperature affects the pressure and the pressure affects the volume afterwards.  Hence to write EoS as,
\be
V = V(P,T)
\ee 
does not make good-enough sense.  We define {\it truly endogenous function}, $\tilde{f}$ which is a function of all independent variables that are the cause of the change of dependent variable. Thus we shall call these independent variables,  {\it truly endogenous variables}.
For hydrostatics system such as gas or solid, truly endogenous function should be expressed as
\be
V = \tilde{V}(P)\,.     
\ee
Writing 
\be
P = \tilde{P}(V)\,,
\ee
is wrong here because $V$ is not the cause of change in $P$.
The other are
\be
P  = \tilde{P}(T)     \;\;\;\;\;\;\;\text{and} \;\;\;\;\;\;\; T = \tilde{T}(P).
\ee
For perfect paramagnetics substance, these are 
\bea
\mathcal{M} & = & \tilde{\mathcal{M}}(B)\,,   \\
\mathcal{M} & = & \tilde{\mathcal{M}}(T)\,, \\
T   & = & \tilde{T}(\mathcal{M})\,.
\eea
Hydrostatics (e.g. gases and solids) and paramagnetics systems are two simple systems of which the many of their EoS are known. Moreover the known EoS of many other systems are similar to either the hydrostatics or the paramagnetics. Systems of elastic string and two-surface thin film are similar to the hydrostatics system due to their mechanical work characters.  The reversible electrolytic cell and dielectric substance are similar to the paramagnetics due to their electromagnetic work characters. Hence, the two systems are good representatives of the two types of EoS.

\section{Effect Diagram for EoS of Hydrostatics and Paramagnetics Systems}
\begin{figure}[h]
\centering
 \includegraphics[width=15pc]{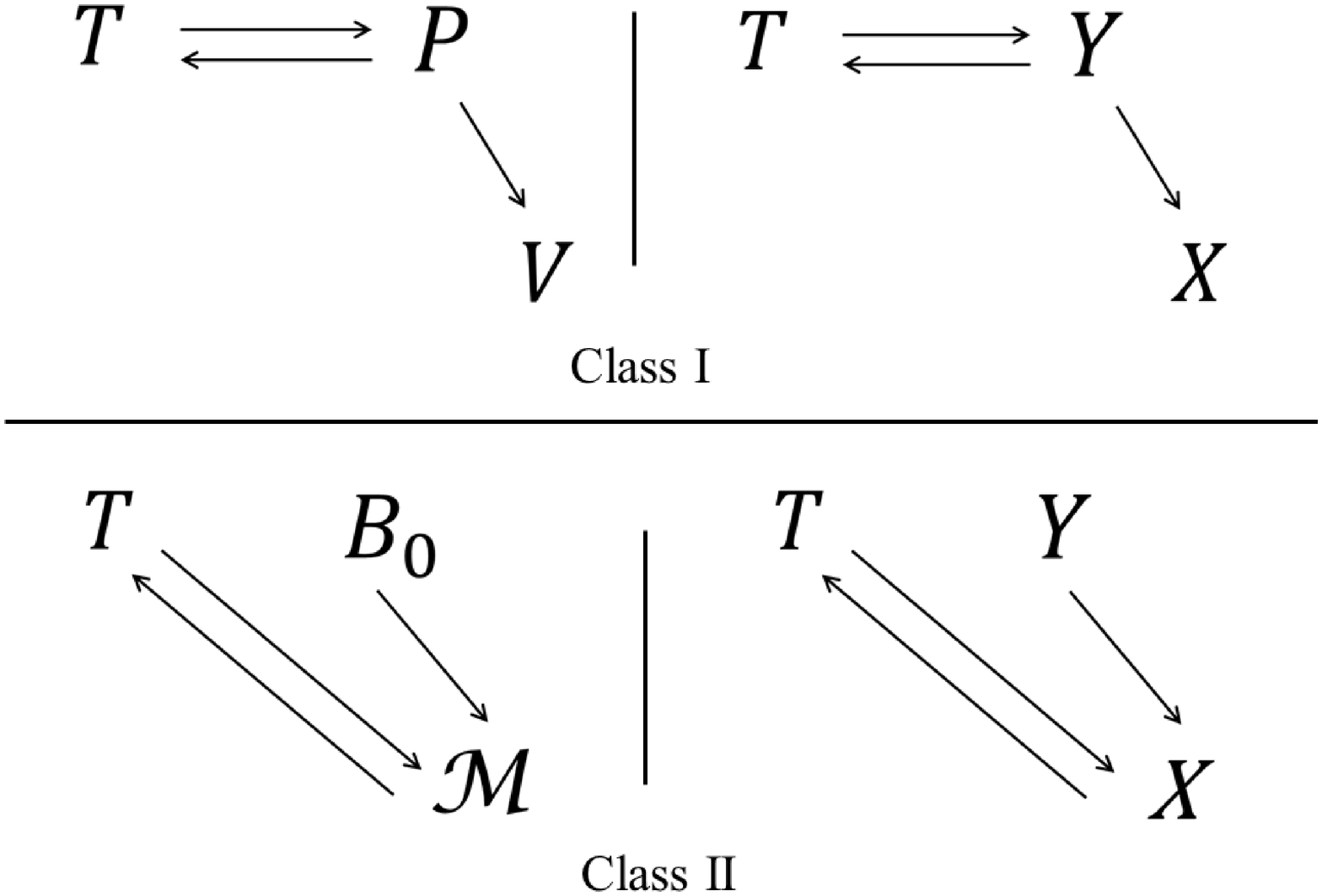}\vspace{0.1pc}\hspace{0.1pc}
\begin{minipage}[b]{22pc}
\caption{\label{figc1} Diagram showing truly endogenous effect in EoS of hydrostatics system (upper) and paramagnetics system (lower). The diagrams represent Class I and Class II of EoS. On the right, to be generalized, diagrams of the two classes are expressed with extensive coordinate $X$ and intensive coordinate $Y$.}
\end{minipage}
\end{figure}

We analyze truly endogenous effect of coordinate variables in the two EoS using arrows pointing from variable that is the cause to the variable that is affected by the cause.
That is to say one variable has active endogenous effect to the other which is done passively. We classify them into two classes as in Fig. \ref{figc1}. 

Expressing ideal gas EoS and perfect paramagnetics EoS (Curie's law) in $X,Y$ variables and let $k$ be a constant magnification factor of combined effects of $Y$ and $T$, we speculate that
\bea
\text{An example of Class I:} \;\;\;\;\;\;\;\;\;\;\;\;    V =  \f{nRT}{P} \;\;\;\;\;\;  & \Longrightarrow & \;\;\;\;\;\; X  = k \l(\f{T}{Y}\r)     \label{cvvv1}   \\
\text{An example of Class II:} \;\;\;\;\;\;\;\;\;\;\;   \mathcal{M} =  C\f{B_0}{T} \;\;\;\;\;\; & \Longrightarrow & \;\;\;\;\;\;  X  = k \l(\f{Y}{T}\r)     \label{cvvv2}
\eea
In ideal gas EoS, $T$ is proportional to $X$, i.e. having enhancing effect to $X$ but $Y$ is inversely proportional to $X$, i.e. having reducing effect to $X$. Situations are reversed in the case of perfect paramagnetics EoS. Mathematical form of Class I and II do not need to be as the equations (\ref{cvvv1})
and (\ref{cvvv2}) which are only examples. Distinct classes are specified with the effect diagram only.

\section{Testing of a Proposed EoS of a Market with Unitary Price Demand and Supply as Linear Function of Price}
\begin{figure}[h]
\centering
 \includegraphics[width=15pc]{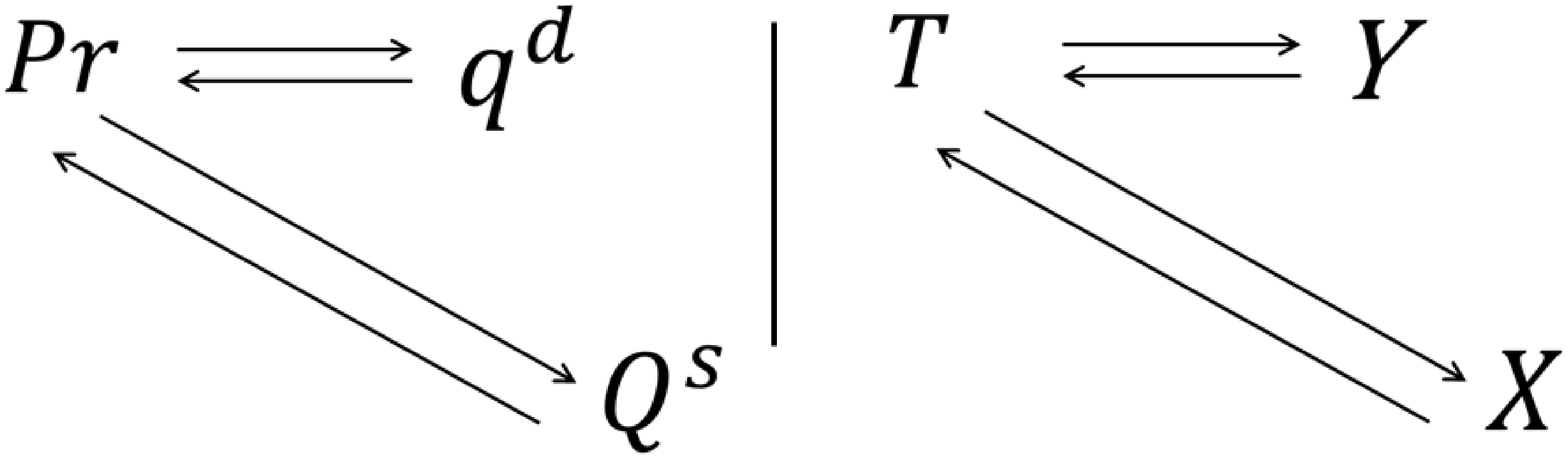}\vspace{0.1pc}\hspace{0.1pc}
\begin{minipage}[b]{22pc}
\caption{\label{figc2} Effect diagram of the proposed EoS of a market with unitary price demand and supply as linear function of price. On the right is the diagram represented with the extensive and intensive coordinates $X$ and $Y$.}
\end{minipage}
\end{figure}

\begin{figure}[h]
\centering
 \includegraphics[width=9pc]{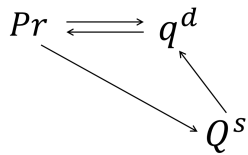}\vspace{0.1pc}\hspace{0.1pc}
\begin{minipage}[b]{22pc}
\caption{\label{figc3} Trying to apply Say's law to the diagram by allowing an arrow from $Q^{\rm s}$ to $q^{\rm d}$ makes a wrong diagram. This is because, in Say's law, the supply affects demand exogenously, not endogenously.}
\end{minipage}
\end{figure}

With aspects we gain and learn from observation of roles and relation between coordinate variables in an EoS as stated above, we will take these aspects as basic criteria for testing of whether an empirical relation could have a status of an EoS. We try EoS of a market with unitary price demand and supply as linear function of price as previously proposed by one of us in Gumjudpai (2018) \cite{GumjMarket}. The proposed EoS, $g(Q^{\rm s}, q^{\rm d}, Pr)   = 0$  is
\be
Q^{\rm s} = \f{1}{K} q^{\rm d} Pr   = 0           \label{propEoS}
\ee
where $Q^{\rm s}$ is supply quantity, $q^{\rm d}$ is demand quantity ($Q^{\rm d}$) per number of consumer ($N$) and $Pr$ is the price of commodity. All are measured at equilibrium  and the market is efficient and information symmetric. The extensive $X$ is $Q^{\rm s}$, the intensive  $Y$ is $q^{\rm d}$.  $T$ is $Pr$ by the assumption that price equilibrium is analogous to thermal equilibrium. That is, temperature identifies thermal equilibrium as of the 0th law of thermodynamics and price identifies the market's price equilibrium as of the 0th law of thermodynamics formulation of economics if it is correct. We draw effect diagram employing known economics facts of price mechanism as in Fig. \ref{figc2}. Written in term of $X, Y$ variables, both $Y$ and $T$ are proportional to $X$.
\be
Q^{\rm s} =  \f{1}{K}{q^{\rm d}}{Pr} \;\;\;\;\;\;  \Longrightarrow \;\;\;\;\;\;  X  = k (Y T)
\ee
We shall see whether this equation should be classified as an EoS and should be of which class. We criticize in two points as follow.
\begin{enumerate}
  \item In Class I and II diagrams, there are three arrows of truly endogenous effect but the diagram in Fig. \ref{figc2}, which is the proposed EoS, there are four arrows. There are two incoming arrows to $Pr$ (analogous to $T$) and two outgoing arrows from $Pr$. We try a specific case of the classical school of economics, that is the Say's law (see textbooks, e.g. \cite{sexton}). The law states that supply quantity $Q^{\rm s}$ dictates demand quantity $Q^{\rm d}$. According to the law, there is no supply surplus left in the market.
      If $Q^{\rm s}$ only affects $Q^{\rm d}$ and it does not affect $Pr$ but instead $Pr$ can affect $Q^{\rm s}$, we hence have diagram as in Fig. \ref{figc3} with four arrows. The consequence is that as $Q^{\rm s}$ increases exogenously, $q^{\rm d}$ increases. As $q^{\rm d}$ increases, the price increases endogenously. Higher price reduces $q^{\rm d}$ endogenously but increases the supply $Q^{\rm s}$ endogenously. As we have higher $Q^{\rm s}$, then $q^{\rm d}$ should be higher. However it does not. This is because the price is now higher and $q^{\rm d}$ is less.
       The law seems to be wrong but it is not the case. We must remember that the effect of $Q^{\rm s}$ to $q^{\rm d}$ according to Say's law must be exogenous, not truly endogenous. How supply can increase demand is because more production results in more income of labour.  Labour people are consumers hence richer consumers have more demand $q^{\rm d}$. This is exogenous effect of the income to $q^{\rm d}$.  Arrows in the context that we propose must be truly endogenous therefore there should not be an arrow from $Q^{\rm s}$ to $q^{\rm d}$ in the diagram. If we want to fix the diagram by removing the arrow from $Q^{\rm s}$ to $q^{\rm d}$, we are left with three arrows as of the Class I and II's diagrams. With three arrows, thus producers then can not influence the price and become price takers. Producing less does not increase the price and this is not true.
  \item  At price equilibrium, $Q^{\rm s} = Q^{\rm d}$. As a result,
  \be
  q^{\rm d}  -  \f{Q^{\rm s}}{N}  = 0\,,  \label{cons}
  \ee
  which is a constraint $h(q^{\rm d}, Q^{\rm s})  =  0$.    Combining with the proposed EoS (\ref{propEoS}), there is only one degree of freedom left for the system. Quasi-static evolution is a process along one-dimensional geometrical object. Hence it is clear that the equation (\ref{propEoS}) constrained with equation (\ref{cons}) is not two-dimensional equilibrium surface but a line.
\end{enumerate}

\section{Conclusion}We consider major features of thermodynamics coordinate variables of various physical systems. We analyze and criticize on roles of these EoS's variables as they are intensive and extensive coordinates and temperature. In doing such, we aim to find a central concept in building of an EoS from other systems such as an economics system and we aim to find some criteria for judging if an empirical equation could take a status of an EoS. We give definition of truly endogenous function here so that we have concrete tool in analyzing EoS.
We propose diagram presenting endogenous effect among these variables so that they can be classified. This set a foundation in constructing and testing of an EoS, of particular interest here, an economics system. We consider thermodynamics approach of economics in distinct ways from other previous works which have initial setting from optimal concepts of variables. Here we take Carath\'{e}odory's axiom which concerns much on the existence of the EoS derived from empirical facts as the beginning step. In this work, we found that the EoS can be classified into two classes. We use the concepts we found here in analyzing the proposed EoS of a market with unitary price demand and linear-in-price supply functions proposed by one of us \cite{GumjMarket}. We found that the proposed EoS is not a new class and is not even an EoS due to the price equilibrium condition which is a serious constraint condition.
The constraint reduces one degree of freedom of the system resulting only one degree of freedom left therefore the proposed equation is not an EoS surface but only a line.

\section*{Acknowledgements}
We thank Chonticha Kritpetch (IF and University of Phayao) for creating the Figures. The work is supported by NIDA Graduate School of Development Economics and ThEP center.

\end{document}